\documentclass[preprint,showpacs,preprintnumbers,amsmath,amssymb,nofootinbib,showkeys,aps]{revtex4}
\usepackage{graphicx}
\usepackage{dcolumn}
\usepackage{bm}
\usepackage{latexsym}
\usepackage{epsfig}
\usepackage{color}
\usepackage{hyperref}
\usepackage{mathrsfs}
\usepackage{bm}

\newcommand{\be}{\begin{equation}}
\newcommand{\ee}{\end{equation}}
\newcommand{\beqq}{\setlength\arraycolsep{2pt}\begin{eqnarray}}
\newcommand{\eeqq}{\vspace{0cm} \end{eqnarray}}
\newcommand{\bea}{\begin{eqnarray}}
\newcommand{\eea}{\end{eqnarray}}
\newcommand{\like}{\mathscr{L}}

\begin{document}

\title{Bayesian correction of $H(z)$ data uncertainties}

\author{J. F. Jesus$^{1,2}$\footnote{jfjesus@itapeva.unesp.br}}
\author{T. M. Greg\'orio$^1$\footnote{tiago.gregorio07@gmail.com}}
\author{F. Andrade-Oliveira$^{3,4}$\footnote{felipe.andrade@linea.gov.br}}
\author{R. Valentim$^5$\footnote{valentim.rodolfo@unifesp.br}}
\author{C. A. O. Matos$^1$\footnote{carlos@itapeva.unesp.br}}

\affiliation{$^1$Universidade Estadual Paulista (Unesp), C\^ampus Experimental de Itapeva, Rua Geraldo Alckmin 519, 18409-010, Vila N. Sra. de F\'atima, Itapeva, SP, Brazil}
\affiliation{$^2$Universidade Estadual Paulista (Unesp), Faculdade de Engenharia, Guaratinguet\'a, Departamento de F\'isica e Qu\'imica, Av. Dr. Ariberto Pereira da Cunha 333, 12516-410, Guaratinguet\'a - SP, Brasil.}
\affiliation{$^3$IFT-UNESP, S\~ao Paulo, SP - 01140-070, Brazil}
\affiliation{$^4$Laborat\'orio Interinstitucional de e-Astronomia - LIneA, Rua General Jos\'e Cristino, 77, Rio de Janeiro, RJ, 20921-400, Brazil}
\affiliation{$^5$Departamento de F\'{\i}sica, Instituto de Ci\^encias Ambientais, Qu\'{\i}micas e Farmac\^euticas - ICAQF, Universidade Federal de S\~ao Paulo (UNIFESP), Unidade Jos\'e Alencar, Rua S\~ao Nicolau No. 210, 09913-030, Diadema, SP, Brazil}


\begin{abstract}
We compile 41 $H(z)$ data from literature and use them to constrain O$\Lambda$CDM and flat $\Lambda$CDM parameters. We show that the available $H(z)$ suffers from uncertainties overestimation and propose a Bayesian method to reduce them. As a result of this method, using $H(z)$ only, we find, in the context of O$\Lambda$CDM, $H_0=69.5\pm2.5\mathrm{\,km\,s^{-1}Mpc^{-1}}$, $\Omega_m=0.242\pm0.036$ and $\Omega_\Lambda=0.68\pm0.14$. In the context of flat $\Lambda$CDM model, we have found $H_0=70.4\pm1.2\mathrm{\,km\,s^{-1}Mpc^{-1}}$ and $\Omega_m=0.256\pm0.014$. This corresponds to an uncertainty reduction of up to 30\% when compared to the uncorrected analysis in both cases.
\end{abstract}

%
\maketitle


\section{\label{introduction} Introduction}

Measurements of the expansion of the Universe are a central subject in the modern cosmology.
In 1998, observations of type Ia supernovae \cite{riess1997,perlmutter} gave strong evidences of a transition epoch between decelerated and accelerated expansion. Those evidences are also consistent with data from Baryon Acoustic Oscillations (BAO) measurements and the Cosmic Microwave Background Anisotropies (CMB).

Among the many viable candidates to explain the cosmic acceleration, the cosmological constant $\Lambda$ explains very well great part of the current observations and it is also the simplest candidate. It gave to the model formed by cosmological constant plus cold dark matter, the $\Lambda$CDM model,  the status of standard model in cosmology. 
 On the other hand, the $\Lambda$ term presents important conceptual problems in its core, e.g., the huge inconsistency of the quantum derived and the cosmological observed values of energy density, the so-called {\it cosmological constant problem} \cite{weinberg89}. Hence, despite of its observational success, the composition and the history of the universe is still a question that needs further investigation. 

Precise measurements of the cosmic expansion may be obtained through the SNe observations. Although they furnish stringent cosmological constraints, they are not directly measuring the expansion rate $H(z)$ but its integral in the line of sight. 
Today, three distinct methods are producing direct measurements of $H(z)$ namely, through differential dating of the cosmic chronometers \cite{Simon05,Stern10,Moresco12,Zhang12,Moresco15,MorescoEtAl16}, BAO techniques \cite{Gazta09,Blake12,Busca12,AndersonEtAl13,Font-Ribera13,Delubac14} and correlation function of luminous red galaxies (LRGs) \cite{Chuang13,Oka13}, which does not rely on the nature of space-time geometry between the observed object and us.

In this work, we treat the $\Lambda$CDM model expansion history as a generative model for the $H(z)$ data \cite{Hogg}.
However, considering a goodness-of-fit criterion, we discuss a possible overestimation in the uncertainty in the current $H(z)$ data and we propose a new generative model to $H(z)$ data, in order to take into account this overestimation.

This article is structured as follows. In Section \ref{dynamics}, we discussed the basic features of the  $\Lambda$CDM model. In section \ref{observational}, we review the $H(z)$ data available on the literature and compile a sample with 41 data.

In Section \ref{analysis}, we discuss the goodness of fit of $\Lambda$CDM with $H(z)$ data and in Section \ref{correction} we discuss a method to treat $H(z)$ uncertainties and apply it to $\Lambda$CDM with spatial curvature. In subsection \ref{analysisflat}, we apply the same method to flat $\Lambda$CDM. In Section \ref{sec-bic} we compare corrected and uncorrected models by using a Bayesian criterion and in Section \ref{comparison} we compare our results with other $H(z)$ analyses. Finally, in Section \ref{conclusion}, we summarize the results.


\section{\label{dynamics}  Cosmic Dynamics of \texorpdfstring{$\Lambda$}{L}CDM Model}

We start by considering the homogeneous and isotropic FRW line element (with $c=1$):
\begin{equation}
\label{ds2}
  ds^2 = dt^2 - a^{2}(t) \left(\frac{dr^2}{1-k r^2} + r^2 d\theta^2 + r^2{\rm sin}^{2}\theta d \phi^2\right),
\end{equation}
where $a$ is the scale factor, $(r,\theta,\phi)$ are comoving coordinates and the spatial curvature parameter $k$ can assume values $-1$, $+1$ or $0$.

In this background, the Einstein Field Equations (EFE) with a cosmological constant are given by 
\begin{eqnarray}
\label{fried1}
    8\pi G \rho= 3 \frac{\dot{a}^2}{a^2} + 3 \frac{k}{a^2} -\Lambda\\
   -8\pi G p = 2\frac{\ddot{a}}{a} + \frac{\dot{a}^2}{a^2} + \frac{k}{a^2} - \Lambda
   \label{fried2}
\end{eqnarray}
where $\rho$ and $p$ are total density and pressure of the cosmological fluid and $\Lambda$ is cosmological constant. We may write the Friedmann equation (\ref{fried1}) in terms of the observable redshift $z$, which relates to scale factor as $a=\frac{a_0}{1+z}$:
\begin{equation}
H^2=\frac{8\pi G(\rho+\rho_\Lambda)}{3} - k(1+z)^2,
\end{equation}
where $\rho_\Lambda=\frac{\Lambda}{8\pi G}$ and $H\equiv\frac{\dot{a}}{a}$ is the expansion rate. The EFE include energy conservation, so we may deduce the continuity equation from Eqs. (\ref{fried1})-(\ref{fried2}):
\begin{equation}
\dot{\rho}_i+3H(\rho_i+p_i)=0,
\end{equation}
where $(\rho_i,p_i)$ stand for each fluid, be it dark matter, baryons, radiation, neutrinos, cosmological constant or anything else that does not exchange energy. For dark matter and baryons, we have $p_i\sim0$, so they evolve with $\rho_i\propto a^{-3}$, cosmological constant has constant $\rho_\Lambda$ and radiation and neutrinos follow $\rho_i\propto a^{-4}$, so they may be neglected in our work, as we are interested in low redshifts (up to $z\sim2$). So, we may write for our components of interest:
\begin{eqnarray}
\rho_m&=&\rho_{m0}(1+z)^3\\
\rho_\Lambda&=&\rho_{\Lambda0}
\end{eqnarray}
where $\rho_m$ stands for dark matter+baryons. So, the Friedmann equation can be written:
\begin{equation}
\left(\frac{H}{H_0}\right)^2=\frac{8\pi G \rho_{m0} (1+z)^3}{3 H_0^2} + \frac{8\pi G \rho_{\Lambda0}}{3 H_0^2} - \frac{k(1+z)^2}{H_0^2}
\end{equation}
and by defining the density parameters $\Omega_i\equiv\frac{\rho_{i0}}{\rho_{c0}}$, where $\rho_{c0}\equiv\frac{3H_0^2}{8\pi G}$ and $\Omega_k\equiv-\frac{k}{a_0^2H_0^2}$, we may write
\begin{equation}
\left(\frac{H}{H_0}\right)^2=\Omega_m (1+z)^3 + \Omega_k (1+z)^2 + \Omega_\Lambda
\end{equation}
from which we deduce the normalization condition $\Omega_m + \Omega_\Lambda + \Omega_k = 1$, or $\Omega_k = 1 - \Omega_m - \Omega_\Lambda$, so we actually have three free parameters on this equation ($\Omega_m,\Omega_\Lambda,H_0$). Finally, we may write for $H(z)$:
\begin{equation}
\label{hzlcdm}
H(z) = H_0\left[\Omega_m (1+z)^3 + (1 - \Omega_m - \Omega_\Lambda)(1+z)^2 + \Omega_\Lambda\right]^\frac{1}{2}
\end{equation}

As usual, we will call this model, where we allow for spatial curvature, O$\Lambda$CDM. The standard, concordance flat $\Lambda$CDM model has $\Omega_k=0$, thus:
\begin{equation}
\label{hzflatlcdm}
H(z) = H_0\left[\Omega_m (1+z)^3 + 1-\Omega_m\right]^\frac{1}{2}
\end{equation}

\section{\label{observational} \texorpdfstring{$H(z)$}{H(z)} data}

Hubble parameter data as function of redshift yields one of the most straightforward cosmological tests because it is inferred from astrophysical observations alone, not depending on any background cosmological models.

At the present time, the most important methods for obtaining $H(z)$ data are\footnote{See Ref. \cite{zt} for a review.} (i) through ``cosmic chronometers", for example, the differential age of galaxies (DAG) \cite{Simon05,Stern10,Moresco12,Zhang12,Moresco15,MorescoEtAl16}, (ii) measurements of peaks of acoustic oscillations of baryons (BAO) \cite{Gazta09,Blake12,Busca12,AndersonEtAl13,Font-Ribera13,Delubac14} and (iii) through correlation function of luminous red galaxies (LRG) \cite{Chuang13,Oka13}.

\begin{figure}[ht]
\centerline{
\epsfig{figure=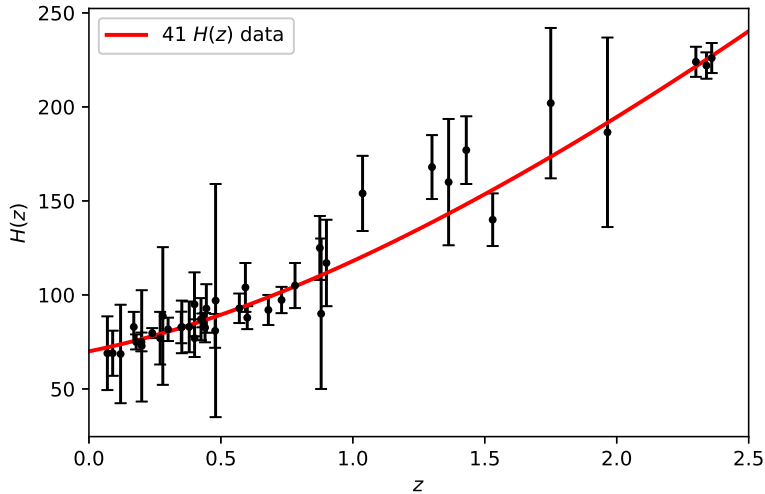,width=.7\linewidth}
}
\caption{41 $H(z)$ data and corresponding best fit $\Lambda$CDM model.}
\label{FigHzData}
\end{figure}

The data we work here are a combination of two compilations: Sharov and Vorontsova \cite{SharovVor14} and Moresco {\it et al.} \cite{MorescoEtAl16}. \cite{SharovVor14} adds 6 $H(z)$ data in comparison to Farooq and Ratra \cite{farooq2013} compilation, which had 28 measurements. Moresco {\it et al.} \cite{MorescoEtAl16}, on their turn, have added 7 new $H(z)$ measurements in comparison to \cite{SharovVor14}. By combining both datasets, we arrive at 41 $H(z)$ data, as can be seen on Table \ref{HzData} and Figure \ref{FigHzData}.

\begin{table}[ht]
\begin{center}
\begin{tabular}{cccc}
\hline\hline
~~$z$ & ~~$H(z)$ &~~~~~~~ $\sigma_{H}$ &~~ Reference\\
\tableline
0.070&~~	69&~~~~~~~	19.6&~~ \cite{Zhang12} \\
0.090&~~	69&~~~~~~~	12&~~	\cite{Simon05} \\
0.120&~~	68.6&~~~~~~~	26.2&~~	\cite{Zhang12} \\
0.170&~~	83&~~~~~~~	8&~~	\cite{Simon05} \\
0.179&~~	75&~~~~~~~	4&~~	\cite{Moresco12} \\
0.199&~~	75&~~~~~~~	5&~~	\cite{Moresco12} \\
0.200&~~	72.9&~~~~~~~	29.6&~~	\cite{Zhang12} \\
0.240&~~	79.69&~~~~~~~	6.65&~~	\cite{Gazta09} \\
0.270&~~	77&~~~~~~~	14&~~	\cite{Simon05} \\
0.280&~~	88.8&~~~~~~~	36.6&~~	\cite{Zhang12} \\
0.300&~~	81.7&~~~~~~~	6.22&~~	\cite{Oka13} \\
0.350&~~	82.7&~~~~~~~	8.4&~~	\cite{Chuang13} \\
0.352&~~	83&~~~~~~~	14&~~	\cite{Moresco12} \\
0.3802&~~	83&~~~~~~~	13.5&~~	\cite{MorescoEtAl16} \\
0.400&~~	95&~~~~~~~	17&~~	\cite{Simon05} \\
0.4004&~~	77&~~~~~~~	10.02&~~\cite{MorescoEtAl16} \\
0.4247&~~	87.1&~~~~~~~	11.2&~~	\cite{MorescoEtAl16} \\
0.430&~~	86.45&~~~~~~~	3.68&~~	\cite{Gazta09} \\
0.440&~~	82.6&~~~~~~~	7.8&~~	\cite{Blake12} \\
0.4497&~~	92.8&~~~~~~~	12.9&~~	\cite{MorescoEtAl16} \\
0.4783&~~	80.9&~~~~~~~	9&~~	\cite{MorescoEtAl16} \\

\hline\hline
\end{tabular}
\quad
\begin{tabular}{cccc}
\hline\hline
~~$z$ & ~~$H(z)$ &~~~~~~~ $\sigma_{H}$ &~~ Reference\\
\tableline
0.480&~~	97&~~~~~~~	62&~~	\cite{Stern10} \\
0.570&~~	92.900&~~~~~~~	7.855&~~\cite{AndersonEtAl13} \\
0.593&~~	104&~~~~~~~	13&~~	\cite{Moresco12} \\
0.6&~~		87.9&~~~~~~~	6.1&~~	\cite{Blake12} \\
0.68&~~		92&~~~~~~~	8&~~	\cite{Moresco12} \\
0.73&~~		97.3&~~~~~~~	7.0&~~	\cite{Blake12} \\
0.781&~~	105&~~~~~~~	12&~~	\cite{Moresco12} \\
0.875&~~	125&~~~~~~~	17&~~	\cite{Moresco12} \\
0.88&~~		90&~~~~~~~	40&~~	\cite{Stern10} \\
0.9&~~		117&~~~~~~~	23&~~	\cite{Simon05} \\
1.037&~~	154&~~~~~~~	20&~~	\cite{Moresco12} \\
1.300&~~	168&~~~~~~~	17&~~	\cite{Simon05} \\
1.363&~~	160&~~~~~~~	22.6&~~	\cite{Moresco15}  \\
1.43&~~		177&~~~~~~~	18&~~	\cite{Simon05} \\
1.53&~~		140&~~~~~~~	14&~~	\cite{Simon05} \\
1.75&~~		202&~~~~~~~	40&~~	\cite{Simon05} \\
1.965&~~	186.5&~~~~~~~	50.4&~~	\cite{Moresco15} \\
2.300&~~	224&~~~~~~~	8&~~	\cite{Busca12} \\
2.34&~~	222&~~~~~~~	7&~~	\cite{Delubac14} \\
2.36&~~	226&~~~~~~~	8&~~	\cite{Font-Ribera13} \\
& & & \\
\hline\hline
\end{tabular}

\end{center}
\caption{41 Hubble parameter versus redshift data.}
\label{HzData}
\end{table}

From these data, we perform a $\chi^2$-statistics, generating the $\chi^2$ function of free parameters:
\begin{equation}
 \chi^2=\sum_{i=1}^{41}\left[\frac{H_0E(z_i,\Omega_m,\Omega_\Lambda)-H_i}{\sigma_{Hi}}\right]^2
 \label{chi2h}
\end{equation}
where $E(z)\equiv\frac{H(z)}{H_0}$ and $H(z)$ is given by Eq. (\ref{hzlcdm}).

\section{\label{analysis} Data analysis and goodness of fit}
In order to minimize the $\chi^2$ function (\ref{chi2h}) and find the constraints over the free parameters $(H_0,\Omega_m,\Omega_\Lambda)$, we have sampled the likelihood $\like\propto e^{-\chi^2/2}$ through Monte Carlo Markov Chain (MCMC) analysis. A simple and powerful MCMC method is the so called Affine Invariant MCMC Ensemble Sampler by Goodman and Weare \cite{GoodWeare}, which was implemented in {\sffamily Python} language with the {\sffamily emcee} software by Foreman-Mackey {\it et al.} \cite{ForemanMackey13}. This MCMC method has the advantage over simple Metropolis-Hasting (MH) methods of depending on only one scale parameter of the proposal distribution and on the number of walkers, while MH methods in general depend on the parameter covariance matrix, that is, it depends on $n(n+1)/2$ tuning parameters, where $n$ is dimension of parameter space. The main idea of the Goodman-Weare affine-invariant sampler is the so called ``stretch move'', where the position (parameter vector in parameter space) of a walker (chain) is determined by the position of the other walkers. Foreman-Mackey {\it et al.} modified this method, in order to make it suitable for parallelization, by splitting the walkers in two groups, then the position of a walker in one group is determined by {\it only} the position of walkers of the other group\footnote{See \cite{AllisonDunkley13} for a comparison among various MCMC sampling techniques.}.

We used the freely available software {\sffamily emcee} to sample from our likelihood in our 3-dimensional parameter space. We have used flat priors over the parameters.
In order to plot all the constraints in the same figure, we have used the freely available software {\sffamily getdist}\footnote{{\sffamily getdist} is part of the great MCMC sampler and CMB power spectrum solver {\sffamily COSMOMC}, by Lewis and Bridle \cite{cosmomc}.}, in its {\sffamily Python} version. The results of our statistical analyses from Eq. (\ref{chi2h}) correspond to the red lines in Fig. \ref{triangle} and Table \ref{Tabf}. From this analysis, we have obtained $\chi^2_\nu=\frac{\chi^2_{min}}{\nu}=18.551/38=0.48819$, where $\nu=n-p$ is number of degrees of freedom.

As it is well known \cite{Bevington,Vuolo}, when one analyses the probability distribution of $\chi^2_\nu$ it has an expected value $\chi^2_\nu = 1 $. $\chi^2_\nu$ values very far from this are unlikely. High $\chi^2_\nu$ values may indicate underestimation of uncertainties or poor fitting of the model, while low values of $\chi^2_\nu$ indicate, in general, overestimation of uncertainties. The $\chi^2_\nu$ distribution is given by
\be
h_\nu(\chi^2_\nu)=\frac{\nu^{\frac{\nu}{2}}(\chi^2_\nu)^{\frac{1}{2}(\nu-2)}e^{-\frac{\nu}{2}\chi^2_\nu}}{2^{\nu/2}\Gamma(\nu/2)},
\label{hnuchi2nu}
\ee
where $\Gamma$ is complete gamma function. It can be shown that the mean $\chi^2_\nu$ is given by $\overline{\chi^2_\nu}=1$, while the mode is given by $\widehat{\chi^2_\nu}=1-\frac{2}{\nu}$. In the limit of a large sample and few parameters, both converge to the same value $\chi^2_\nu \approx 1$. From (\ref{hnuchi2nu}), we may also define the cumulative distribution function (cdf) or probability of obtaining a value of $\chi^2_\nu$ as low as $Q$ as:
\be
P(\chi^2_\nu<Q)\equiv\int_0^{Q} h_\nu(Q')dQ'
\label{cdfchi2nu}
\ee

\begin{figure}[ht]
\centerline{
\epsfig{figure=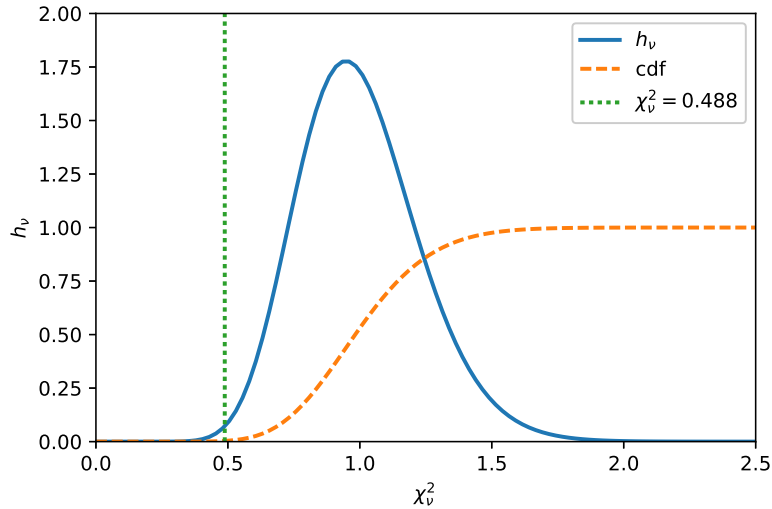,width=.7\linewidth}
}
\caption{$h_\nu(\chi^2_\nu)$ and corresponding cdf for $\nu=38$.}
\label{probchi2red}
\end{figure}

In order to realize how low is the $\chi^2_\nu$ value we have obtained, namely, $\chi^2_\nu=0.48819$, we have plotted the pdf $h_\nu(\chi^2_\nu)$ (\ref{hnuchi2nu}) and the cdf (\ref{cdfchi2nu}) for $\nu=38$ in Fig. \ref{probchi2red}.

As one may see in this figure, the probability of obtaining $\chi^2_\nu$ as low as $\chi^2_\nu=0.488$ for $\nu=38$ is quite low. In fact, by calculating the integral (\ref{cdfchi2nu}), we have obtained $P(\chi^2_\nu<0.48819)=0.3342\%$. It indicates, thus, a very low and unlikely $\chi^2$ value, which, in turn, from Eq. (\ref{chi2h}) indicates overestimated $H(z)$ uncertainties.


\section{\label{correction}\texorpdfstring{$H(z)$}{H(z)} uncertainties correction}
How one may try to correct uncertainties? Ideally, at the level of obtaining data, new methods less prune to errors are to be used. In fact, in general, data coming from BAO and Lyman $\alpha$ have smaller errors than data coming from differential ages. However, not being able to reobtaining the data, or reanalyzing then through new methods, we are left with the available data. Then, nothing can be done? From the Bayesian viewpoint, not necessarily. In fact, we may view the data as a collection of $(z_i,H_i,\sigma_{Hi})$. Very often, we are interested in a likelihood given by $\mathcal{L}=Ne^{-\chi^2/2}$, where $N$ is only a normalization constant and one is interested in maximize the likelihood, which is equivalent to minimize the $\chi^2$. Let us recall from where this expression comes from.

As explained in \cite{Hogg}, the likelihood may be seen as an {\it objective function}, that is, a function that represents monotonically the quality of the fit. Given a scientific problem at hand, as fitting a model to the data, one must define this objective function that represents this ``goodness of fit'', then try to optimize it in order to determine the best free parameters of the model that describe the data.

Hogg {\it et al.} \cite{Hogg} argues that the only choice of the objective function that is truly justified -- in the sense that it leads to probabilistic inference, is to make a {\it generative model} for the data. We may think of the generative model as a parameterized statistical procedure to reasonably generate the given data.

For instance, assuming Gaussian uncertainties in one dimension, we may create the following generative model: Imagine that the data really come from a function $y=f(x,\theta)$ given by the model, and that the only reason that any data point deviates from this model is that to each of the true $y$ values a small $y$-direction offset has been added, where
that offset was drawn from a Gaussian distribution of zero mean and known variance $\sigma_y^2$.  In this model, given an independent position $x_i$, an uncertainty $\sigma_{yi}$, and free parameters $\theta$, the frequency distribution $p(y_i|x_i,\sigma_{yi},\theta)$ for $y_i$ is
\begin{equation}\label{eq:objectivei}
p(y_i|x_i,\sigma_{yi},\theta) = \frac{1}{(2\pi)^{1/2}\sigma_{yi}} \,\exp\left[-\frac{(y_i - f(x_i,\theta))^2}{2\,\sigma_{yi}^2}\right] \quad ,
\end{equation}

Thus, if the data points are independently drawn, the likelihood $\like$ is the product of conditional probabilities
\begin{equation}\label{eq:like}
\like = \prod_{i=1}^n \ p(y_i|x_i,\sigma_{yi},\theta) \quad .
\end{equation}
Taking the logarithm,
\begin{equation}
\ln\like = -\frac{1}{2} \sum_{i=1}^n\left[ \frac{(y_i - f(x_i,\theta))^2}{\sigma_{yi}^2}+\ln(2\pi\sigma_{yi}^2)\right]
 \label{lnlike}
\end{equation}

In equation above, the second term $-\frac{1}{2}\sum_i\ln(2\pi\sigma_{yi}^2)$ is in general absorbed in the likelihood normalization constant, because the variances $\sigma_{yi}^2$ are considered fixed by the data. Here, we consider $\sigma_i$ as parameters to be obtained by optimization of the objective function $\like$. As discussed in \cite{Hogg}, it can be considered a correct procedure from the Bayesian point of view, although an involved one, and the obtained $\sigma_i$ can be quite prior dependent.

In order to avoid having more free parameters than data, here we consider the $\sigma_i$ to be all overestimated by a constant factor $f$, thus, $\sigma_{i,true}=f\sigma_i$. It can be seen just as a simplifying hypothesis, a first order correction. More elaborated methods could be cluster the data in some groups, then correct the $\sigma_i$ for each group. However, as explained in \cite{Hogg}, it is not an easy task to separate good data from bad data, and not necessarily the bad data are the ones with bigger uncertainties. So, we limit ourselves here with just one overall correction factor, next we conclude if this a good approximation. We treat $f$ as a free parameter, then we constrain it in a joint analysis with the cosmological parameters, similar to what is made in current SNe Ia analyses \cite{Union2,Union2.1,Betoule}. For $\Lambda$CDM, then, our set of free parameters now is $\theta=(H_0,\Omega_m,\Omega_\Lambda,f)$. A simpler but less justified hypothesis would be simply find the value for $f$ which provides $\chi^2_\nu\equiv1$. However, as we expect $\chi^2_\nu$ to have some variance, such a procedure is not much trustworthy. With $f$ as a free parameter, it may include some uncertainty into the analysis, when compared to the standard, uncorrected analysis, but at the same time, it may also reduce the cosmological parameters uncertainties.

Instead of Eq. (\ref{lnlike}), we must work here with the following objective function:
\begin{equation}
\ln\like = -\frac{1}{2} \sum_{i=1}^n\left\{\frac{\left[H_i - H(z_i,H_0,\Omega_m,\Omega_\Lambda)\right]^2}{f^2\sigma_{Hi}^2}+\ln(2\pi f^2\sigma_{Hi}^2)\right\}
 \label{lnlikef}
\end{equation}

\begin{figure}[ht]
\centerline{
\epsfig{figure=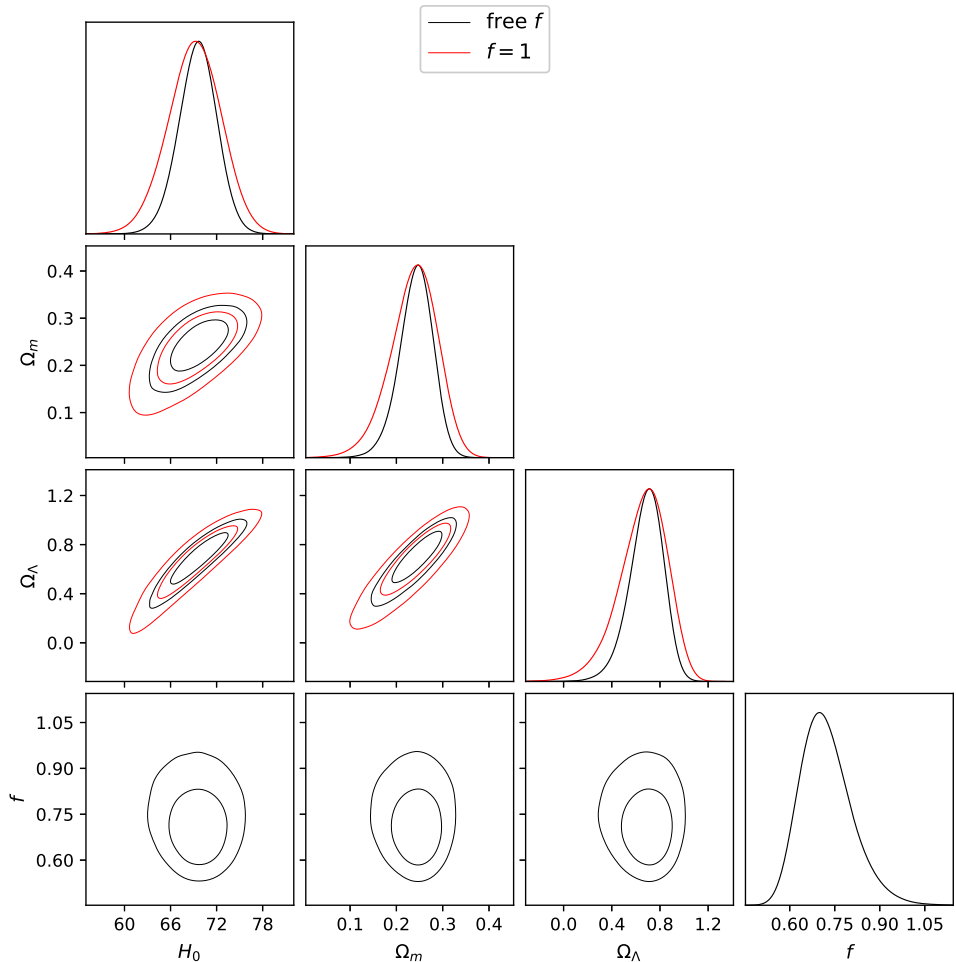,width=.8\linewidth}
}
\caption{The results of statistical analysis for O$\Lambda$CDM model. $H_0$ in km/s/Mpc. {\bf Diagonal:} Marginalized constraints from $H(z)$ data for each parameter. {\bf Below diagonal:} Marginalized contours constraints for each indicated combination of parameters, with contours for 68.3\% and 95.4\% confidence levels.}
\label{triangle}
\end{figure}

By maximizing the above likelihood, we find not only the best fit cosmological parameters, but also the best correction factor $f$ which will furnish the best model to describe the data. By doing the same procedure of last section, now with the additional parameter $f$, we find the constraints shown by the black lines on Figure \ref{triangle}.

From Figure \ref{triangle}, we may already see the difference in the parameter space if we introduce the $f$ parameter. The corrected contours (black lines) are narrower then the uncorrected contours (red lines). It can be quantified by the parameter constraints shown on Table \ref{Tabf}.

\begin{table}[h]
\begin{tabular} { c  c c c c}
\hline
                 & $H(z)$ only     &                           & $H(z)+H_0$               & \\
\hline
 Parameter       &  Uncorrected    & Corrected                 & Uncorrected              & Corrected\\
\hline
$H_0$            & $69.1\pm 3.5  $ & $69.5\pm 2.5$             & $72.4\pm1.5$             & $72.5\pm1.1$          \\

$\Omega_m$       & $0.237\pm0.051$ & $0.242\pm0.036$           & $0.267\pm0.038$          & $0.268\pm0.028$\\

$\Omega_\Lambda$ & $0.66\pm0.20  $ & $0.68\pm0.14$             & $0.825^{+0.11}_{-0.095}$ & $0.831\pm0.073$\\

$f$              & --              & $0.723^{+0.084}_{-0.085}$ & --                       & $0.728^{+0.067}_{-0.098}$\\
\hline
\end{tabular}
\caption{Mean values of parameters of O$\Lambda$CDM model from $H(z)$ data, without uncertainties correction and with uncertainties correction factor $f$. Uncertainties correspond to 68\% c.l.\label{Tabf}}
\end{table}

As can be seen on Table \ref{Tabf}, $\sigma_{H0}$ has been reduced from 3.5 to 2.5, $\sigma_{\Omega_m}$ has been reduced from 0.051 to 0.036 and $\sigma_{\Omega_\Lambda}$ has been reduced from 0.20 to 0.14. The mean value for $f$ was $f=0.723^{+0.084}_{-0.085}$. An interesting feature we may see from Fig. \ref{triangle}, is that the $f$ parameter is much uncorrelated to cosmological parameters. It explains the small shift on mean values of cosmological parameters from Table \ref{Tabf}. Saying in another way, the central values of cosmological parameters are insensitive to overall shifts on $H_i$ uncertainties, but their variances are directly affected by $f$.

\subsection{\label{analysisflat}Flat \texorpdfstring{$\Lambda$}{L}CDM}
For completeness, as flat $\Lambda$CDM model is favoured from many observations, in this section we analyse this model similarly to O$\Lambda$CDM. Eq. (\ref{hzlcdm}) now reads:
\be
H(z) = H_0\left[\Omega_m (1+z)^3  + 1-\Omega_m\right]^\frac{1}{2}
\ee

\begin{figure}[ht]
\centerline{
\epsfig{figure=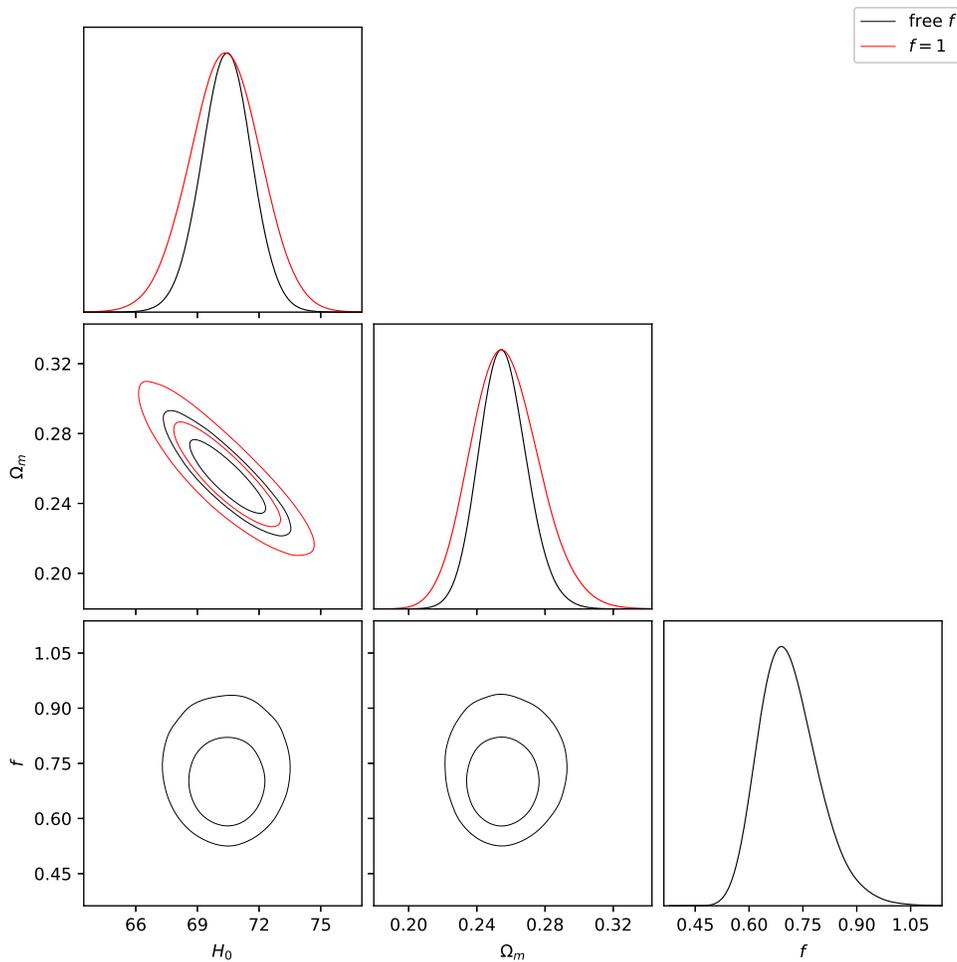,width=.8\linewidth}
}
\caption{The results of statistical analysis for flat $\Lambda$CDM model. $H_0$ in km/s/Mpc. {\bf Diagonal:} Marginalized constraints from $H(z)$ data for each parameter. {\bf Below diagonal:} Marginalized contours constraints for each indicated combination of parameters, with contours for 68.3\% and 95.4\% confidence levels.}
\label{triangleFlat}
\end{figure}

The results of this analysis may be seen on Fig. \ref{triangleFlat} and Table \ref{TabFlat}.

As one may see from Fig. \ref{triangleFlat}, $f$ is again uncorrelated to cosmological parameters, so it does not change their central values.

\begin{table}[h]
\begin{tabular} { c  c c c c}
\hline
           & $H(z)$ only     &                 & $H(z)+H_0$                &                           \\
\hline
 Parameter &  Uncorrected    & Corrected       & Uncorrected               & Corrected                 \\
\hline
$H_0$      & $70.3\pm 1.7  $ & $70.4\pm 1.2$   & $71.8\pm1.2$              & $71.80\pm0.89$            \\

$\Omega_m$ & $0.257\pm0.020$ & $0.256\pm0.014$ & $0.243^{+0.014}_{-0.015}$ & $0.242\pm0.011$           \\

$f$        & --              & $0.714\pm0.082$ & --                        & $0.728^{+0.066}_{-0.096}$ \\
\hline
\end{tabular}
\caption{Mean values of parameters of Flat $\Lambda$CDM model from $H(z)$ data, without uncertainties correction and with uncertainties correction factor $f$. Uncertainties correspond to 68\% c.l.\label{TabFlat}}
\end{table}

As one may see on Table \ref{TabFlat}, the $H_0$ uncertainty, for instance, is reduced from 1.7 to 1.2, which now corresponds to 1.7\% relative uncertainty. $\Omega_m$ uncertainty has reduced from 0.020 to 0.014.

\section{Bayesian Criterion Comparison\label{sec-bic}}
Here, we use the Bayesian Information Criterion (BIC) \cite{Schwarz78,bic-ccdm} in order to compare the models with uncertainties corrections and without uncertainties correction. As an approximation for the Bayesian Evidence (BE) \cite{Trotta08}, BIC is useful because it is, in general, easier to calculate. BIC is given by:
\be
\mathrm{BIC}=-2\ln\like_{max}+p\ln n
\label{bic}
\ee
where $\like_{max}$ is the likelihood maximum and $p$ is the number of free parameters. The two models we want to compare are: $M_1: f=1$, that is, $\Lambda$CDM model without uncertainties correction is enough to describe the data; and $M_2:f\neq1$ such that some correction $f$ to uncertainties is necessary in order to $\Lambda$CDM explain the $H(z)$ data. We may write the log-likelihood as:
\be
\ln\like=-\frac{1}{2}\left[\frac{\chi^2}{f^2}+\sum_{i=1}^n\ln(2\pi f^2\sigma_i^2)\right]
\label{loglike}
\ee
where $\chi^2$ is the uncorrected $\chi^2\equiv\sum_{i=1}^n\frac{\left[H_i - H(z_i,H_0,\Omega_m,\Omega_\Lambda)\right]^2}{\sigma_{Hi}^2}$. To calculate BIC, we must find the maximum of $\ln\like$. By deriving (\ref{loglike}) with respect to $f$:
\be
\frac{\partial\ln\like}{\partial f}=-\frac{1}{f}\left[n-\frac{\chi^2}{f^2}\right]
\ee
When it vanishes, we find the best fit:
\be
\hat{f}=\sqrt{\frac{\chi^2_{min}}{n}}
\label{bestf}
\ee
From (\ref{bic}) and (\ref{bestf}) we find:
\be
\mathrm{BIC}_1=\chi^2_{min}+\sum_{i=1}^n\ln(2\pi\sigma_i^2)+p_1\ln n
\ee
\be
\mathrm{BIC}_2=n+n\ln\left(\frac{2\pi\chi^2_{min}}{n}\right)+\sum_{i=1}^n\ln(\sigma_i^2)+p_2\ln n
\ee
where $p_j$ is the number of free parameters in $M_j$. So:
\be
\Delta\mathrm{BIC}=\mathrm{BIC}_1-\mathrm{BIC}_2=\chi^2_{min}-n\ln\left(\chi^2_{min}\right)+(n-p_2+p_1)\ln n-n
\ee
For $p_1=3$ and $p_2=4$, it simplifies to:
\be
\Delta\mathrm{BIC} = \chi^2_{min} - n\ln\left(\chi^2_{min}\right) + (n-1)\ln n-n
\ee
For $n=41$ and $\chi^2_{min}=18.551$, it yields: $\Delta\mathrm{BIC}=6.352$. As discussed in \cite{bic-ccdm}, for example, values of $\Delta\mathrm{BIC}>5$ corresponds to a decisive or strong statistical difference. That is, by this criterion, the model $M_1$ (no correction) may be discarded against model $M_2$ (with correction).

\section{\label{comparison}Comparison with other \texorpdfstring{$H(z)$}{H(z)} data analyses}
Farooq and Ratra \cite{farooq2013} have constrained O$\Lambda$CDM model with 28 $H(z)$ data and two possible priors over $H_0$. With the most stringent prior, namely, the one from Riess {\it et al.} (2011) \cite{Riess11}, they have found, at 2$\sigma$, $0.20\leq\Omega_m\leq0.44$ and $0.62\leq\Omega_\Lambda\leq1.14$. We have found $0.13\leq\Omega_m\leq0.34$ and $0.23\leq\Omega_\Lambda\leq1.04$ for 41 $H(z)$ data without correction and $0.162\leq\Omega_m\leq0.31$ and $0.38\leq\Omega_\Lambda\leq0.96$ with the $f$ correction. By considering the prior from Riess {\it et al.} (2011), namely, $H_0=73.8\pm2.4\mathrm{\,km\,s^{-1}\,Mpc^{-1}}$, we have found $0.18\leq\Omega_m\leq0.34$ and $0.57\leq\Omega_\Lambda\leq1.04$ without correction and $0.21\leq\Omega_m\leq0.32$ and $0.65\leq\Omega_\Lambda\leq0.99$ with the $f$ correction.

With 34 $H(z)$ data, Sharov and Vorontsova \cite{SharovVor14} find a more stringent result, namely, $H_0=70.26\pm0.32$, $\Omega_m=0.276^{+0.009}_{-0.008}$ and $\Omega_\Lambda=0.769\pm0.029$. However, they have combined $H(z)$ data with SNe Ia and BAO data, which is beyond the scope of our present work. However, by comparing their result with our Table \ref{Tabf}, we may see that both constraints are compatible at 1$\sigma$ c.l.

Moresco {\it et al.} have used their compilation of 30 $H(z)$ data combined with $H_0$ from Riess {\it et al.} (2011) \cite{Riess11} to constrain the transition redshift from deceleration to acceleration, in the context of O$\Lambda$CDM \cite{zt}:
\begin{equation}
 z_t=\left[\frac{2\Omega_\Lambda}{\Omega_m}\right]^{1/3}-1
\end{equation}

They have found $z_t=0.64^{+0.11}_{-0.07}$. By using the present 41 $H(z)$ data, we find $z_t=0.77\pm0.22$ without correction and $z_t=0.78\pm0.15$ with the $f$ correction. The results are in fully agreement without the correction and are compatible at 2$\sigma$ c.l. with the $f$ correction. We have mentioned the mean value for $z_t$, while Moresco {\it et al.} refers to the best fit value.


The constraints over $H_0$ are quite stringent today from many observations \cite{RiessEtAl16,Planck16}. However, there is some tension among $H_0$ values estimated from different observations \cite{BernalEtAl16}, so we choose not to use $H_0$ in our main results here, Figs. \ref{triangle} and \ref{triangleFlat}. We combine $H(z)+H_0$ only in Tables \ref{Tabf} and \ref{TabFlat} and in the present section, using Riess {\it et al.} (2011) \cite{Riess11} result, in order to compare with other earlier analyses.

\section{\label{conclusion}Conclusion}
In this work, we have compiled 41 $H(z)$ data and proposed a new method to better constrain models using $H(z)$ data alone, namely, by reducing overestimated uncertainties through a Bayesian approach. The Bayesian Information Criterion was used to show the need for correcting $H(z)$ data uncertainties. The uncertainties in the parameters were quite reduced when compared with methods of parameter estimation without correction and we have obtained an estimate of an overall correction factor in the context of O$\Lambda$CDM and flat $\Lambda$CDM models.

Further investigations may include constraining other cosmological models or trying to optimally group $H(z)$ data and then correcting uncertainties.

\begin{acknowledgments}
J. F. Jesus is supported by Funda\c{c}\~ao de Amparo \`a Pesquisa do Estado de S\~ao Paulo - FAPESP (Processes no. 2013/26258-4 and 2017/05859-0). FAO is supported by Coordena\c{c}\~ao de Aperfei\c{c}oamento de Pessoal de N\'ivel Superior - CAPES. TMG is supported by Unesp (Pr\'o Talentos grant), R. Valentim is supported by Funda\c{c}\~ao de Amparo \`a Pesquisa do Estado de S\~ao Paulo - FAPESP (Processes no. 2013/26258-4 and 2016/09831-0).
\end{acknowledgments}

\end{document}